# Link between chips and cutting moments evolution

Olivier CAHUC [1,a], Claudiu BISU [2,b] and Alain GERARD [1,c*]

[1] University of Bordeaux and CNRS UMR 5295, I2MI Département MPI,
351 cours de la Libération 33405 Talence FRANCE

[2] University Politehnica of Bucharest, Laboratory Machines and Systems of Production,
Spl. Idependentei, no.313. Bucharest, Romania

[a]olivier.cahuc@u-bordeaux1.fr , [b]claudiu.bisu@upb.ro, [c*]ajr.gerard@gmail.com

**Abstract.** The better understanding of the material cutting process has been shown with the benefit of the forces and moments measurement since some years ago. In paper, simultaneous six mechanical components and chip orientation measurements were realized during turning tests. During these tests, the influence of the depth of cut or feed rate has been observed and a link between the chip orientation and the moment vector orientation or the central axis characteristics has been shown.

## Nomenclature

| | |
|---|---|
| **BT** | Block Tool |
| **BW** | Block Workpiece |
| ap | Depth of cut (mm) |
| $D_1$ | Holding fixture diameter (mm) |
| $D_2$ | Workpiece diameter (mm) |
| E | Young modulus (N/mm$^2$) |
| $F_i$ | Force components vector (i = x, y, z) (N) |
| f | Feed rate (mm) |
| $L_1$ | Holding fixture length (mm) |
| $L_2$ | Length workpiece (mm) |
| $M_i$ | Torque components vector (i = x, y, z) (dN. N) |
| MyCop | Torque component along y axis direction in insert frame (dN.m) |
| MzCop | Torque component along z axis direction in insert frame (dN.m) |
| $\|\|M\|\|$ | Torque modulus (dN.m) |
| **WTM** | Workpiece-Tool-Machine |
| x (z) | Cross (feed) direction |
| y | Cutting axis |
| $\gamma$ | Cutting angle (°) |
| $\lambda_s$ | Inclination angle of edge (°) |
| $\alpha$ | Clearance angle (°) |
| $\kappa_r$ | Direct angle (°) |
| $r_\varepsilon$ | Nozzle radius (mm) |
| R | Sharpness radius (mm) |
| $\theta$ | Chip ejection angle (°) |

* corresponding author

**Introduction**

Metal cutting is one of the most important manufacturing processes. The most common cutting processes are turning, milling, drilling and grinding.

During the cutting process of different materials, a whole of physic-chemical and dynamic phenomena are involved. Elasto-plastic strains, friction and thermal phenomena are generated in the contact zone between workpiece, tool and chip. These phenomena are influenced by: physical properties of the materials (workpiece and tool), tool geometry, cutting and lubrication conditions, and also the machining system dynamic parameters (stiffness, damping) [13]. The machine tool vibrations are generated by the interaction between the elastic machining system and the cutting process. The elastic system is composed of: the different parts of the machine tool in movement, the workpiece and the tool. Actions of the machining process are usually forces and moments. These actions also generate relative displacements of elements composing the elastic system. They occur for example between the tool and workpiece, the tool device and bed, etc. These displacements modify the cutting conditions and in the same way the mechanical actions. Thus, the knowledge of the machining system elastic behavior is essential to understand the cutting process [8].

In these processes, the cutting forces measurement has important and tremendous applications within industry and research alike. The cutting forces estimation allows to supervise tool wear evolution [16], establishes material machinabilities, optimizes cutting parameters, predicts machined workpiece surface quality and study phenomena such as chip formation or vibrations appearance.

Knowing the cutting forces is essential to machine tool builders in calculating power requirements and frame rigidity [1]. Cutting forces acting on the tool must be measured at the design tool that are strong enough to remove chip at the desired quantity from the workpiece and to calculate power of tool driver system. The dynamometer is able to measure three force components: cutting force, feed force and radial force but not the torque at the tool tip. Axinte et al., [2] propose a procedure to account for both calibration and process errors in the uncertainty estimation for the specific situation of single cutting force measurements.

However the dynamometer can measure three perpendicular cutting force components and three torque components simultaneously during turning, and the measured numerical values can be stored in computer by data acquisition system [10]. This dynamometer was designed to measure up to 5,000 N maximum force and 350 dN/m torque. The system sensitivity is ± 4% in force and ± 6% in torque.

**Experimental device**

Dynamic cutting tests are carried out on a lathe Ramo (RTN30) for which the spindle speed does not exceed 6,000 rpm (Fig. 1). The main components of the test machining system used are presented without its data-processing environment in the Figure 2. Moreover, a six-components dynamometer [9], being used as tool-holder [18], is positioned on the lathe to measure all the cutting mechanical actions (forces and torques).

The three-dimensional dynamic character is highlighted by seeking the various existing correlations between the various parts of the machining system and the various parameters evolutions, which ensure to characterize the process.

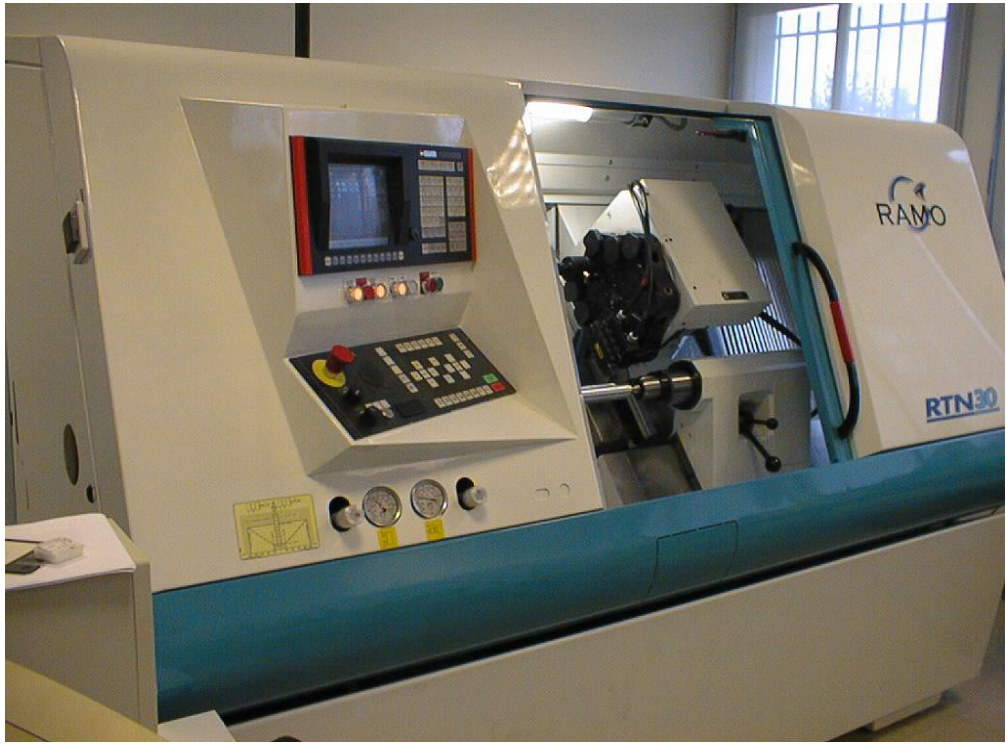

Fig 1 Lathe Ramo.

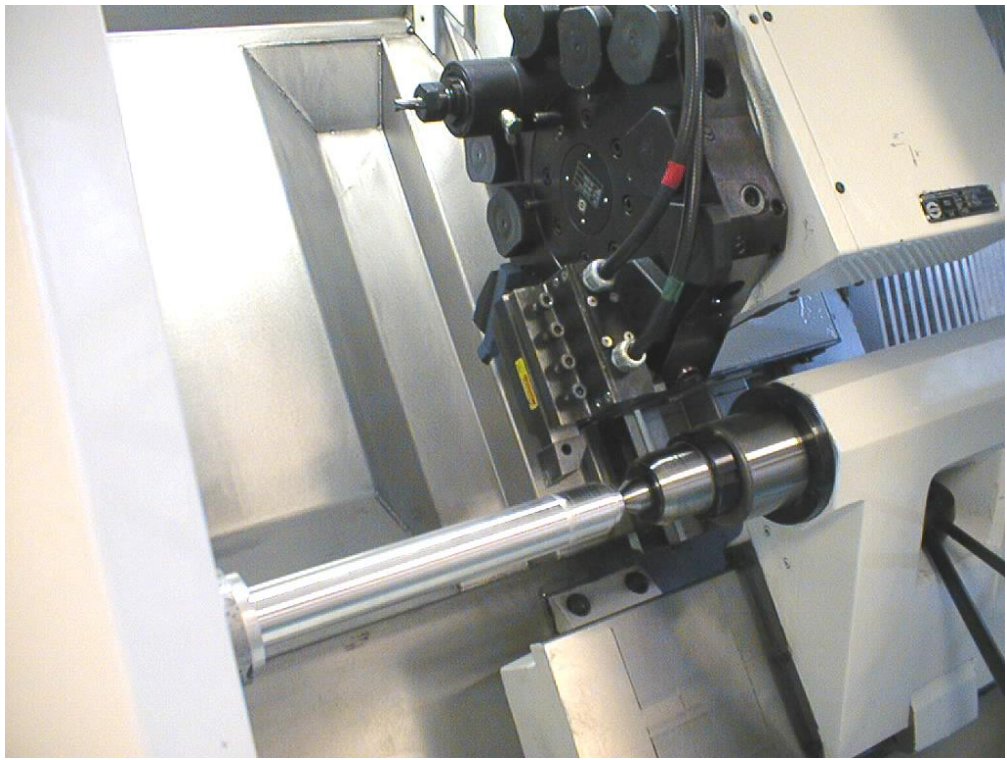

Fig 2 Experimental machining system and metrology environment in turning process.

The **BT** part includes the tool, the tool-holder, the dynamometer, the fixing plate on the cross slide. The six-component dynamometer [10] is fixed between the cross slide and the tool-holder (Fig. 2). This is necessary thereafter to measure the cutting mechanical actions.

An example of the measurement is given in the Fig. 3 and Fig. 4.

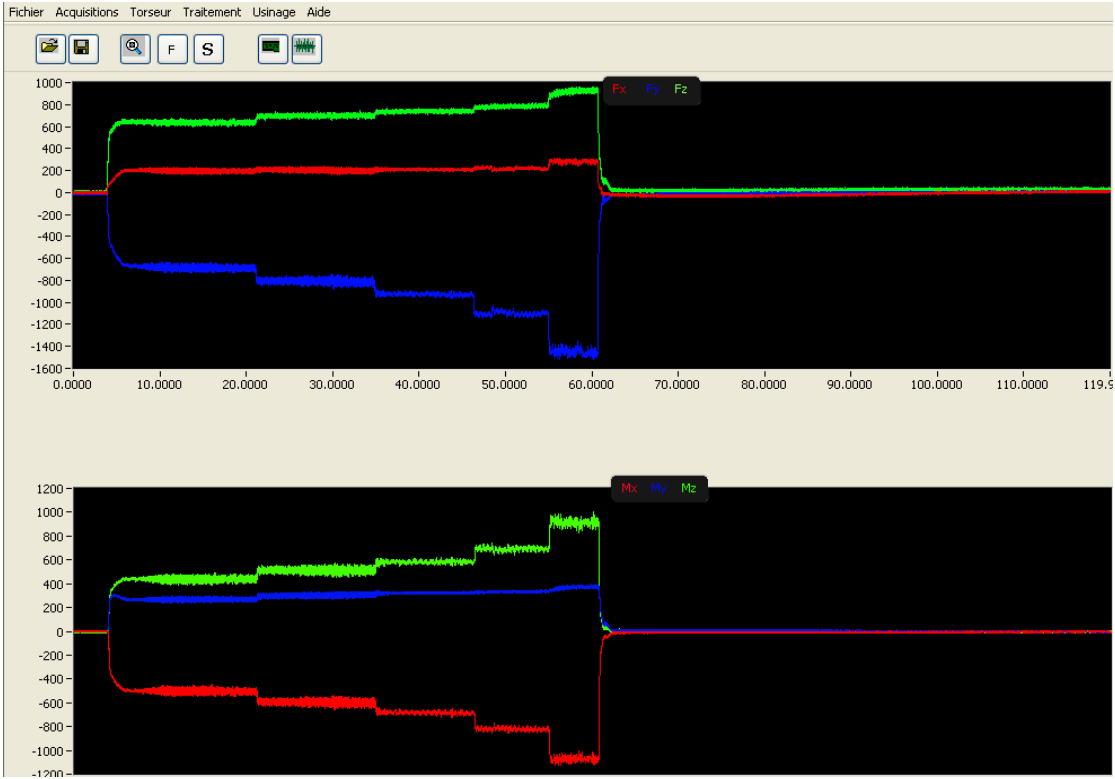

Fig. 3 Signals related to the force and moment components following the three **x**, **y**, **z** cutting directions (Fig. 2); test case using parameters ap = 3.5 mm, f variable and Vc = 238 m/min.

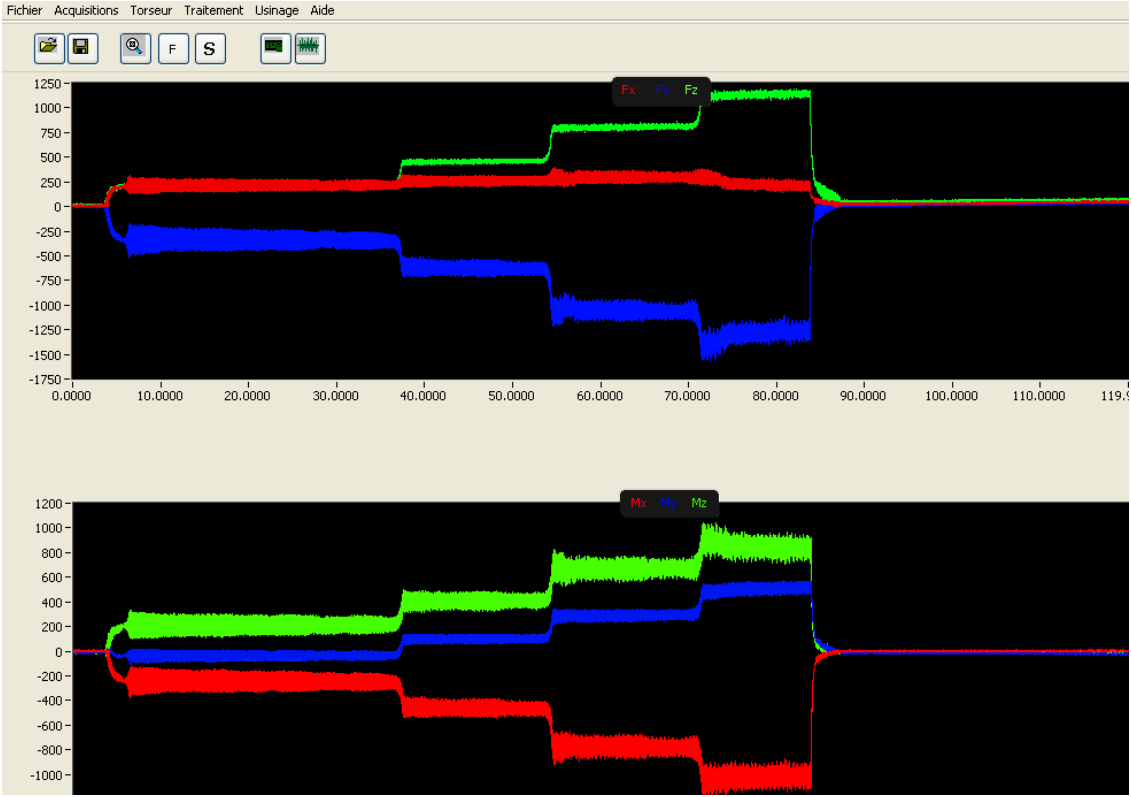

Fig. 4 Signals related to the force and moment components following the three **x**, **y**, **z** cutting directions (Fig. 2); test case using parameters ap variable, f = 0.1 mm/rev and Vc = 238 m/min.

## Tests results

For each test, except the feed rate or the depths of cut values, all the turning parameters are constant. The mechanical actions are measured according to the feed rate (f) using the six components dynamometer following the method developed and finalized by Couétard [10] and used in several occasions [7, 11, 12, 15, 16]. On the experimental device, (Fig. 2) the instantaneous spindle speed is permanently controlled.

During the tests the insert tool used is type TNMA 16 04 08 carbide not covered, without chip breaker. The machined material is a steel alloy of chrome molybdenum type 42CD4T. The test-workpieces are cylindrical with a diameter of 116 mm (Fig. 5 and Fig. 6). They were designed starting from the Finite Elements Method being coupled to a procedure of optimization described in [7].

Moreover, the tool geometry is characterized by the cutting angle $\gamma$, the clearance angle $\alpha$, the inclination angle of edge $\lambda_s$, the direct angle $\kappa_r$, the nozzle radius $r_\varepsilon$ and the sharpness radius R [14]. The tool parameters are detailed in the Table 1.

Table 1: Geometrical characteristics of the tool

| $\gamma$ | $\alpha$ | $\lambda_s$ | $\kappa_r$ | $r_\varepsilon$ | R |
|---|---|---|---|---|---|
| -6° | 6° | -6° | 91° | 0.8 mm | 1.2 mm |

## Cutting torsor actions

**Tests.** The experiments are performed within a framework similar to the one described in Cahuc et al., [7]. For each test, the complete torsor of the mechanical actions are measured using the six-component dynamometer.

These mechanical actions are evaluated in two different configurations. The first one with constant feed rate f= 0.1 mm/rev (for ap = 3.5 mm), variable depths of cut: ap = 1; 2; 3.5; 5 mm (for f = 0.1 mm) and the test-workpieces is cylindrical with a diameter of 116 mm (Fig. 5 and 6).

For every value of ap the manufacturing is made with landings 20 mm in length on a cylinder beforehand prepare in staircase (Fig. 5) The second one has a different depths of cut: ap = 3.5 mm and variable feed rate f : f = 0.05; 0.0625; 0.075; 0.1; 0.15 mm/rev on the same initial diameter of 116 mm. The diameter of the cylindrical test tube is constant and every manufacturing is made on a 20 mm length (Fig. 6).

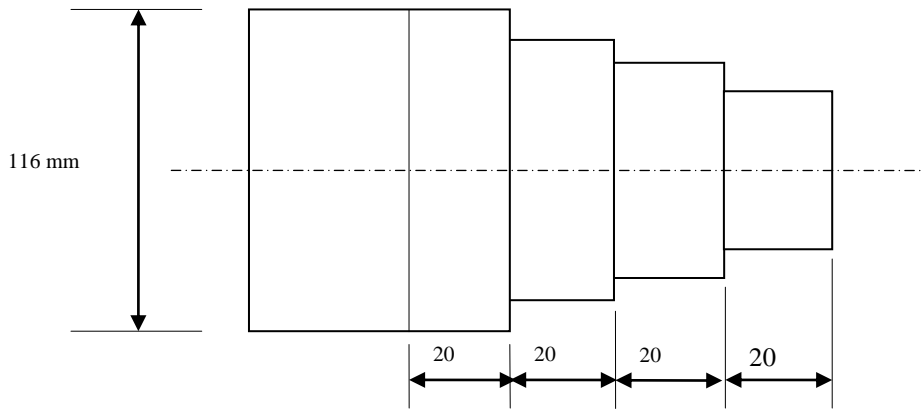

Fig 5 : Workpiece geometry for variable ap and for feed rate f=0.1 mm/rev

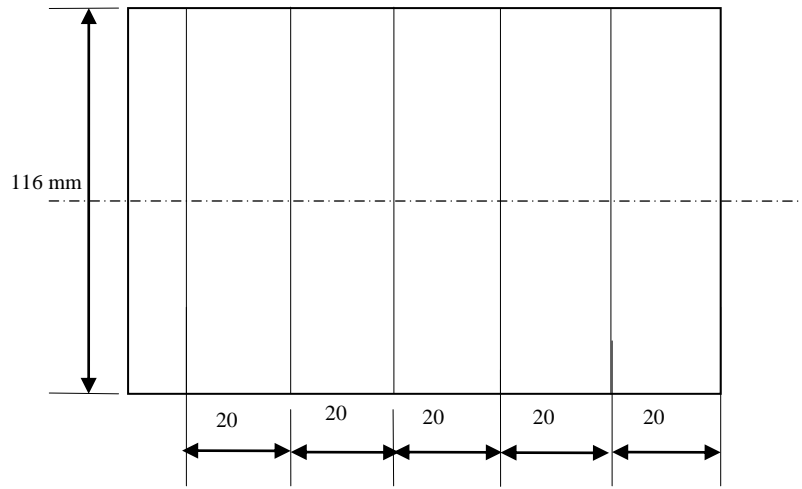

Fig 6 : Workpiece geometry for ap=3.5 mm and f variable

The six-component dynamometer gives the instantaneous values of all the torque cutting components in the three-dimensional space (**x, y, z**) related to the machine tool (Fig. 2). Measurements are performed in O', which is the center of the six-component dynamometer. Then, they are transported to the tool point O via the moment transport classical relations (Eq. 4). Measurement uncertainties of the six-component dynamometer are: ± 4 % for the force components and ± 6 % for the moment components.

**Force components of the cutting actions analysis.** For the first test feed rate f = 0.1 mm and four depth of cut indicated above, an example of average torsor (force, moment) measurements applied to the tool tip point are presented on Table 2 (where θ is the chip ejection angle).

Table 2: Average torsor components measured at the tool tip point (f=0.1 mm/rev)

| $\theta$ (°) | ap (mm) | $F_x$ (N) | $F_y$ (N) | $F_z$ (N) | $\|R\|$ (N) | $M_x$ (dNm) | $M_y$ (dNm) | $M_z$ (dNm) | $M_{yCop}$ (dNm) | $M_{zCop}$ (dNm) | $\|M\|$ (dNm) |
|---|---|---|---|---|---|---|---|---|---|---|---|
| 36 | 1 | 213 | -341 | 225 | 461 | 11 | -31 | -28 | -0,8 | -4,1 | 43.7 |
| 51 | 2 | 258 | -630 | 456 | 819 | 5 | -23 | -28 | 0,7 | -3,5 | 36.3 |
| 79 | 3.5 | 294 | -1,060 | 805 | 1,363 | 9 | -44 | -62 | 5,9 | -5,5 | 76.7 |
| 88 | 5 | 226 | -1,287 | 1,136 | 1,731 | 67 | -104 | -40 | 3,8 | -10,5 | 103.1 |

For the five values feed rate f indicated above, one example of torsor (force, moment) measurements applied to the tool tip point is presented on Table 3.

Table 3: Average torsor components measured at the tool tip point (ap=3.5 mm)

| $\theta$ (°) | f (m/rev) | $F_x$ (N) | $F_y$ (N) | $F_z$ (N) | $\|R\|$ (N) | $M_x$ (dNm) | $M_y$ (dNm) | $M_z$ (dNm) | $M_{yCop}$ (dNm) | $M_{zCop}$ (dNm) | $\|M\|$ (dNm) |
|---|---|---|---|---|---|---|---|---|---|---|---|
| 78 | 0.05 | 198 | -983 | 637 | 435,087 | 5.2 | - 25.8 | -34 | 2,8 | -3,2 | 43 |
| 70 | 0.0625 | 202 | -808 | 698 | 564,385 | 4.6 | - 28.7 | - 45.1 | 3,2 | -4,2 | 53.7 |
| 79 | 0.075 | 209 | -926 | 739 | 684,155 | 6.1 | - 33.3 | - 57.3 | 5 | -4,4 | 66.6 |
| 83 | 0.1 | 217 | -1,099 | 784 | 862,076 | 2.2 | - 47.4 | - 68.7 | 6,2 | -5,6 | 83.6 |
| 69 | 0.15 | 281 | - | 922 | 1,338,822 | 9.8 | -5 6.9 | - 90.7 | 6,4 | -8,6 | 107,6 |

The analysis of Table 2 and Table 3 confirms the following results obtained in [4]:

|Fx| < |Fz| < |Fy|     (1)

|$M_{oz}$| < |$M_{oy}$| < |$M_{ox}$| , (whatever f, in the machine frame )     (2)

We notice that the absolute value of the force components is increasing with ap that is not the case for the moment's components.
On the other hand, in the frame tool, the relation (3) is always valid about is f or ap inrease.

|$M_{yCop}$| < |$M_{zCop}$| < |$M_x$| , (whatever f or ap, in the tool frame )     (3)

We also remark that the absolute value of the moment component in Y direction (or Z) are always increasing whereas the component on X direction passes by an extremum in the tool frame.
As we wish to look at the sensibility of moments with diverse parameters, we restrict our investigations on the moment's evolution.

## Moments analysis at central axis

**Central axis.** It is well-known that, it is possible to associate a central axis, to any torsor (except the torsor of pure moment), which is the single object calculated starting from the six torsor components [6]. A torsor $[A]_0$ in a point O is composed of the resultant forces **R** and the resulting moment $M_0$.

$$[A]_O = \begin{cases} R \\ M_o \end{cases} \quad (4)$$

The central axis is the straight line classically defined by:

$$\mathbf{OA} = \frac{\vec{R} \wedge \vec{M_o}}{\|R^2\|} + \lambda \vec{R} \quad (5)$$

where O is the point where the mechanical actions torsor was moved (here, the tool tip) and A the current point describing the central axis. Thus, **OA** is the vector associated with the bi-point [O, A] (Fig. 7).
This line (Figure 7 a) corresponds to the geometric points where the mechanical actions moment torsor is minimal. The central axis calculation consists in determining the points assembly (a line) where the torsor can be expressed along a slide block (straight line direction) and the pure moment (or torque) [6].

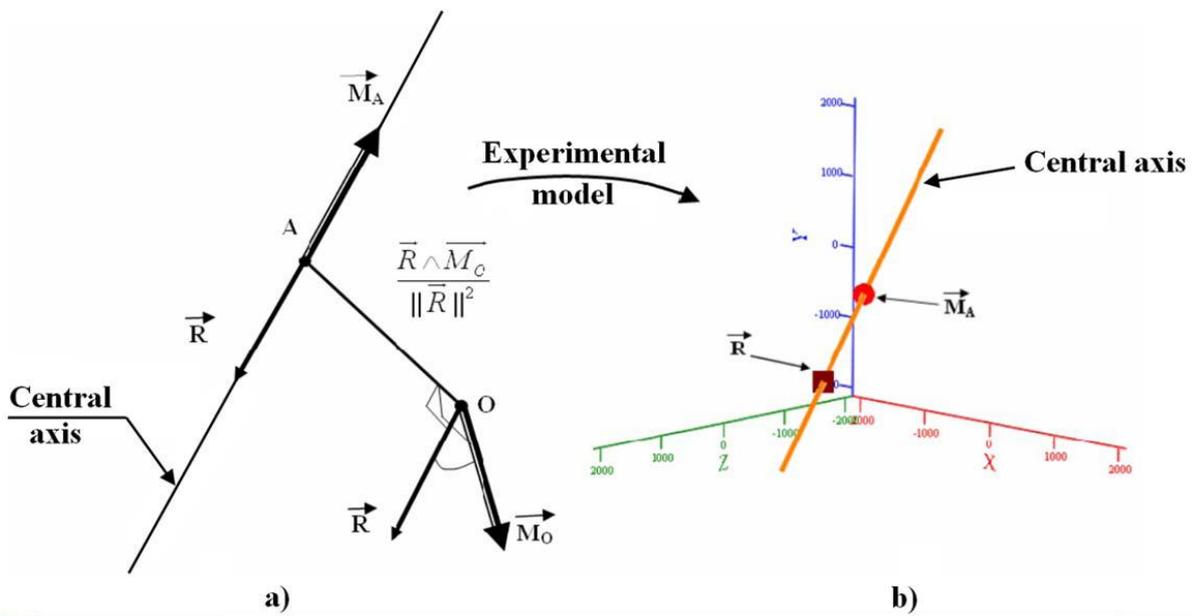

Fig. 7 Central axis representation (a) and of the colinearity between vector sum **R** and minimum moment **M**$_A$ on central axis (b)

**Central axes beams analysis.** The results given to the previous paragraphs allow us to determine the central axes beams. Figure 8 gives a representation of these in the case f = 0.1 mm / rev (ap = 3.5 mm) and ap variable (for f = 0.1 mm/rev.). In addition, the Figure 9 gives the evolution of the central axes in the case ap = 3.5 mm and variable f.

In both cases, we notice the convergence of all central axes towards the same very narrow zone of elliptic shape (a quasi point) placed in a plan in agreement with [3]. This zone corresponds instead of the summits of the cones of the central axes beams, which evolve in time. This evolution is led by the light vibrations of the system manufacturing that we can notice on the represented examples of metrology Fig. 3 and Fig. 4. Besides, considering the definition of the central axes, the narrow zone of convergence of these central axes is the place of the points where moments are minimum. However, the principle of the virtual works shows that vectorials spaces efforts and movements are two dual spaces. In other words, the properties established for a vectorial space are it also for its dual space. As a result, the properties of minimum revealing for moments are also valid for the displacements. Thus, it is the center of the rigidity (minimum place of the movements) of the system, which we so determine by another method than, that used in [3].

Furthermore we notice that when f increases, the angles of the axes of the cones of the central axes beams with regard to the axis Z increase (Fig. 10) while it decreases when ap increase (Fig. 11).

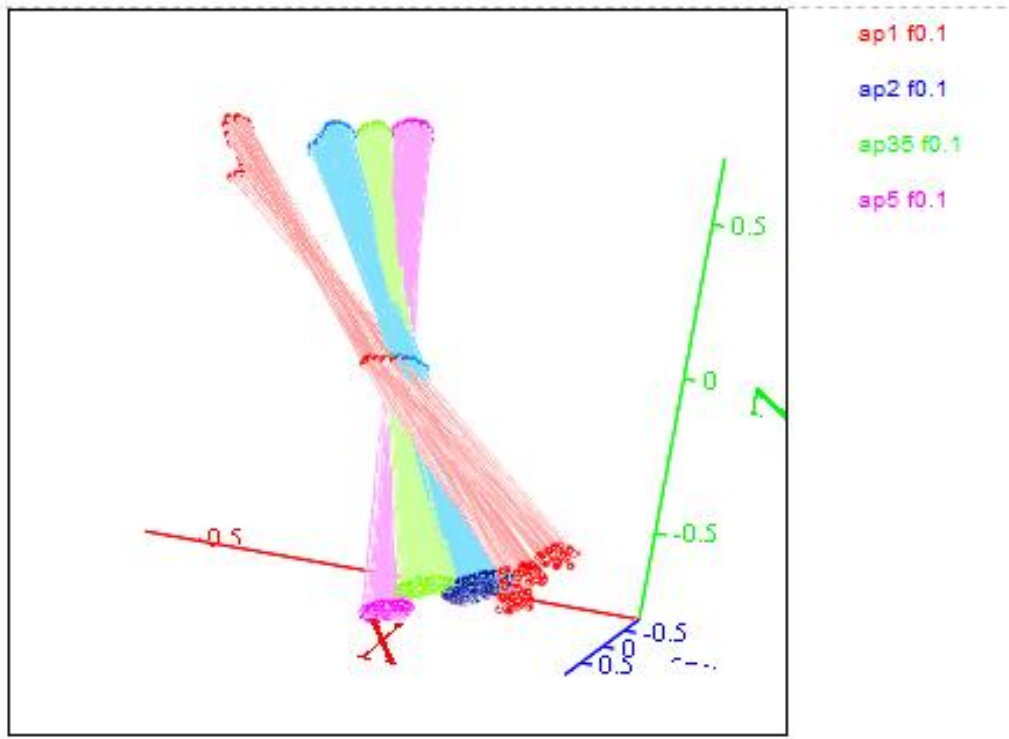

Fig. 8 Central axes beams for f=0.1, ap variable

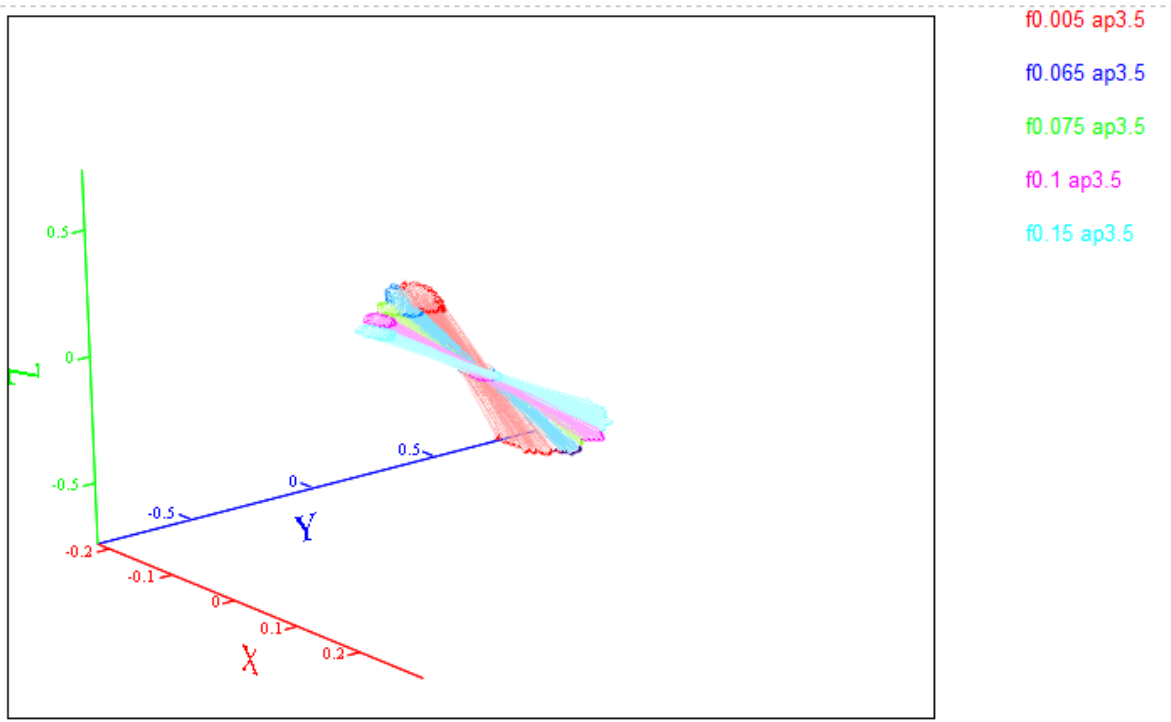

Fig. 9 Central axes beams for ap =3.5 mm, f variable

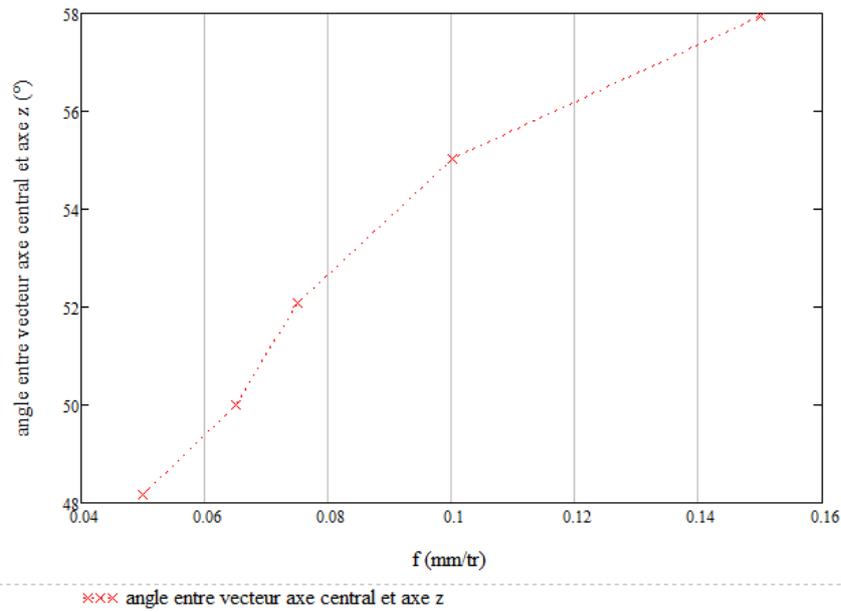

Fig. 10 Angle of the axes of the cones of the central axes with the axis Z when f increases.

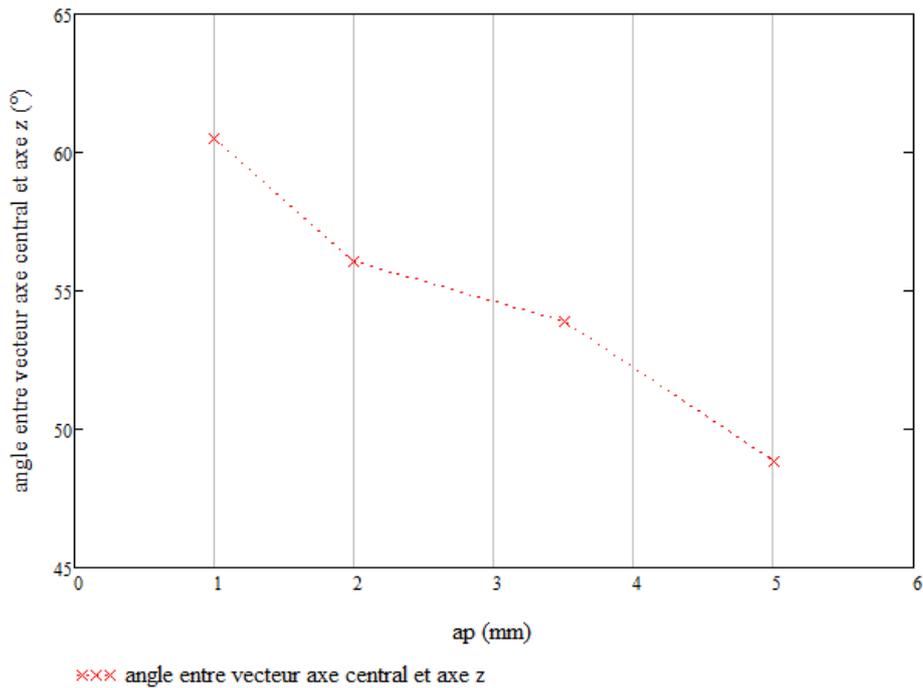

Fig. 11 Angle of the axes of the cones of the central axes with the axis Z when ap increases.

In the case of the increasing evolution of ap for f=0.1 mm / rev we observe that the angular variation of the central axes with the axis Z is quasi-linear (Fig. 11). On the other hand when f grows for ap=3.5 mm the evolution of the angle of the central axes with the axis Z is almost a curve of degree 2 (Fig. 10).

The evolution of the module of moments reported to the chip ejection angle when f or ap increase is also interesting.

**Moment modulus analysis.** The module of the moments is proportional to ap (resp. f) and θ when for f = 0.1 mm/rev. ap (resp. for ap = 3.5 mm f) increases (Fig. 12). For these tests, we see that the chip ejection angle module has a linear evolution of the moment with θ when ap variable (type θ = 0.95 IIMII/θ + 0.01). This lightly parabolic evolution of chip ejection angle is not very clear when look at the moment of module when f increases.

On the other hand, this quasi-parabolic evolution of the chip ejection angle function of the module of the moment when ap increases is confirmed (Fig. 13). But, the chip ejection angle evolution function of the module of the moment when f increases is not so clear, it seems parabolic. This allows us to seize better the differences observed in the behavior of the central axes beams as ap (Fig. 8) or f (Fig. 9) is increasing.

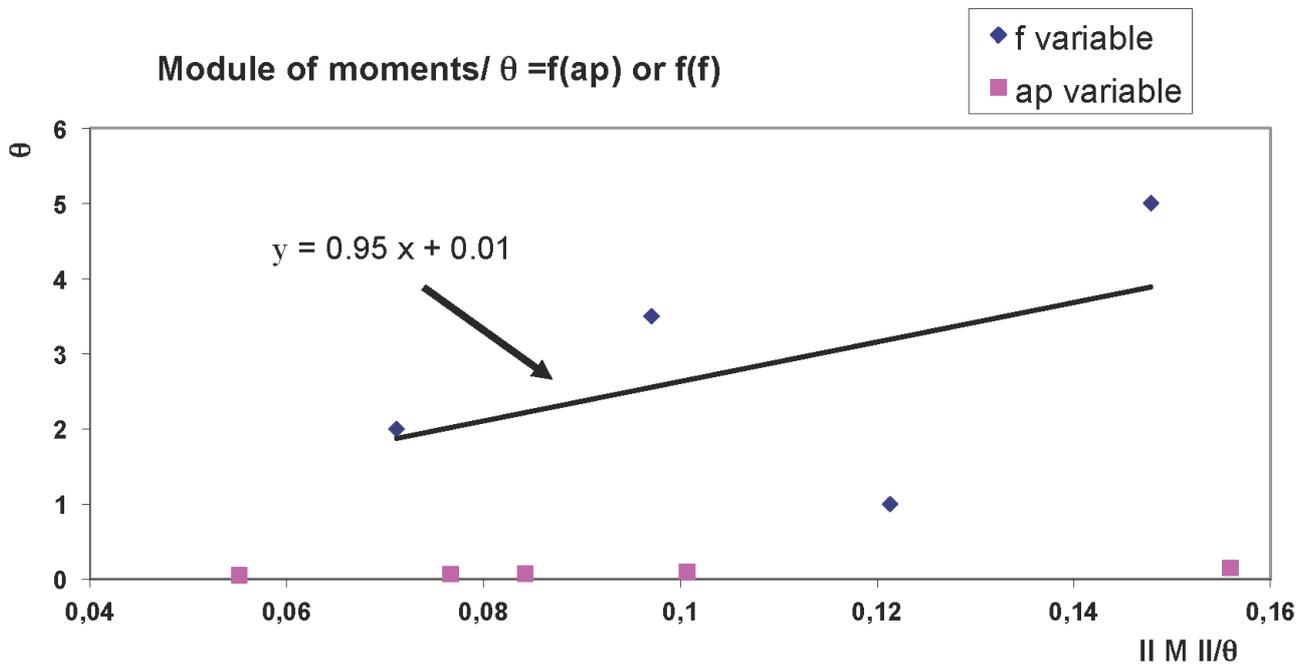

Fig. 12 Evolution of the module of moments for ap = 3.5 mm and variable f

It is also interesting to look at the evolution of the chip ejection angle function of the moments component in the frame tool when ap varies (Fig. 14) and when f varies (Fig. 15).

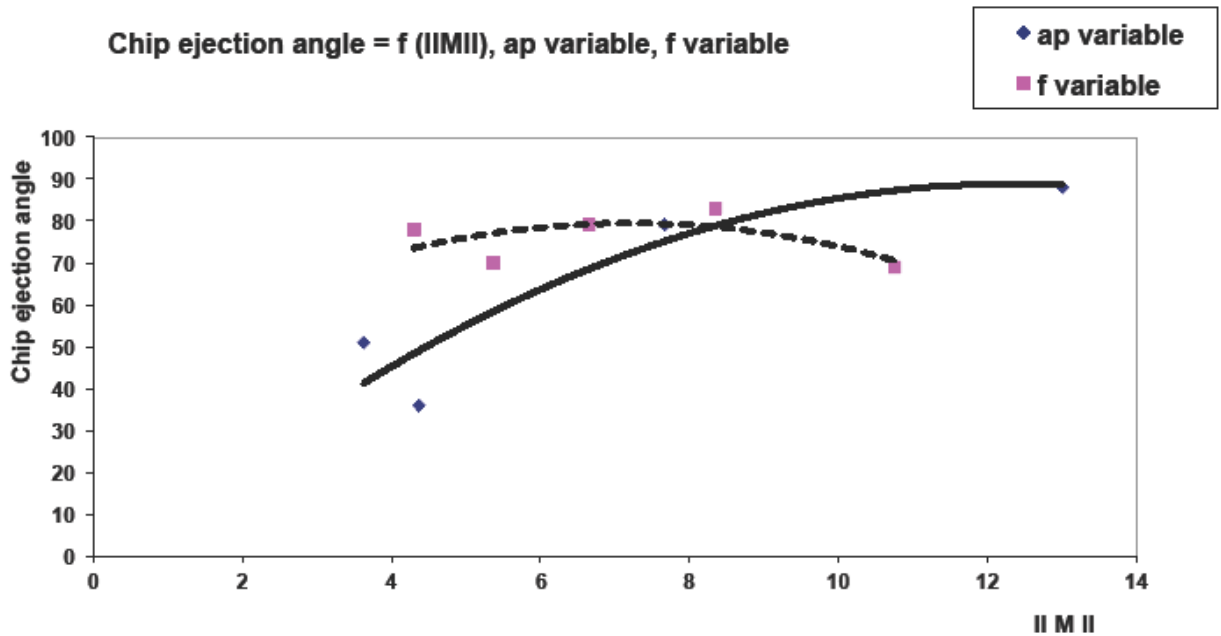

Fig; 13 Chip angle evolution function of the moment modulus with ap and f increase.

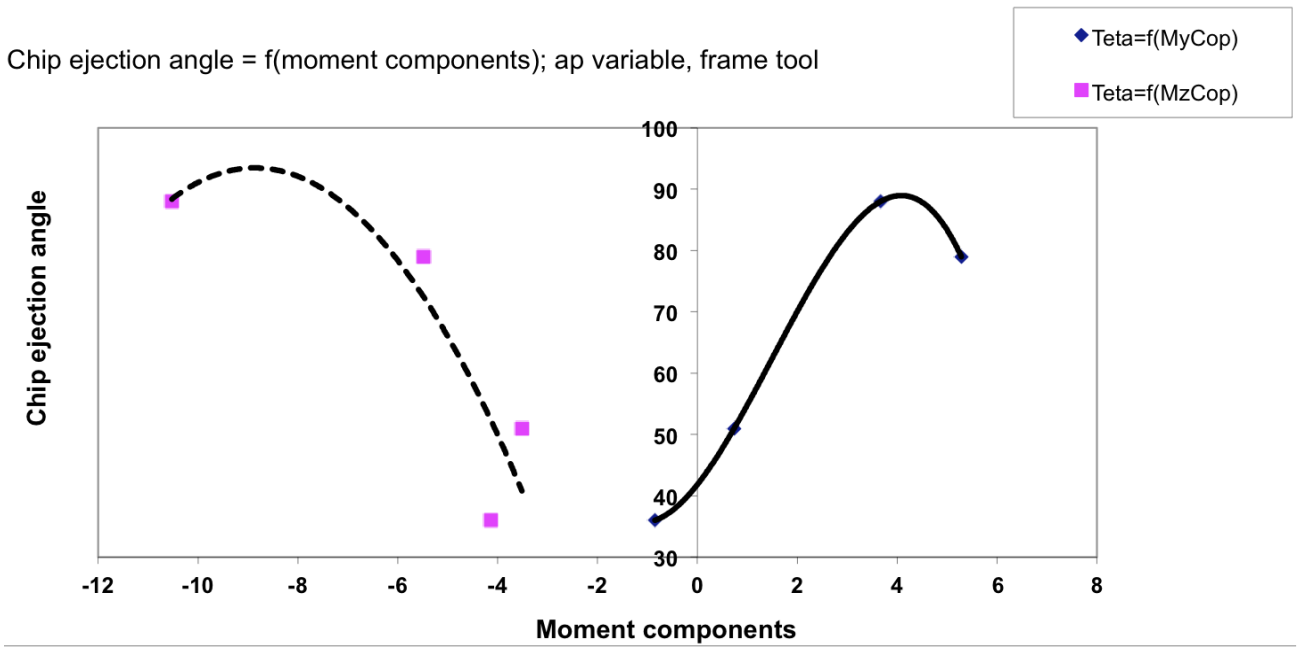

Fig. 14 Chip angle evolution function of the moment components when ap increases (f = 0.1 mm.rev.).

When ap increases (for f = 0.1 mm/rev; Fig. 14) MzCop has a parabolic evolution and seems to have and an asymptote whereas MyCop has only a maximum.

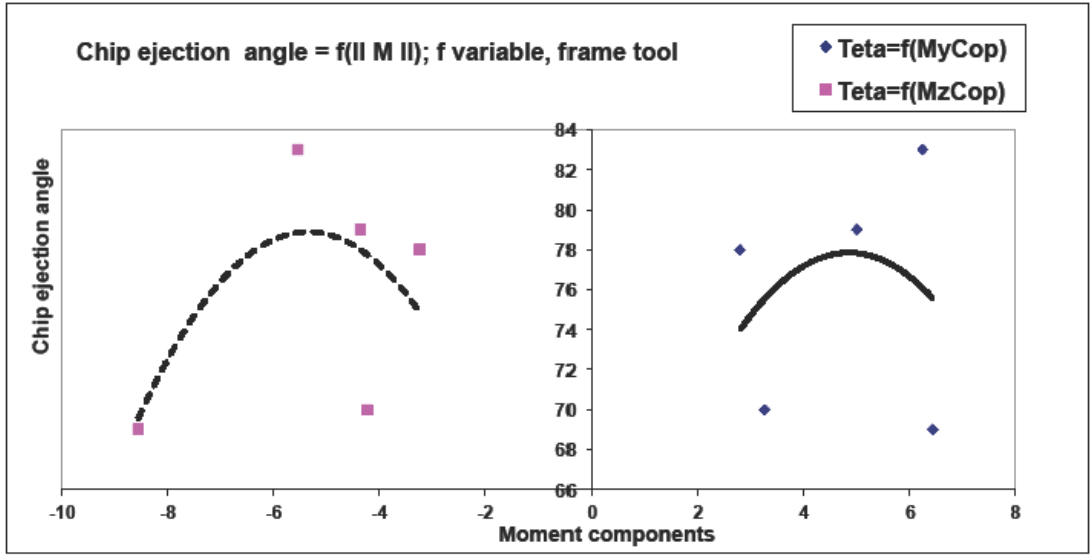

Fig. 15 Evolution of the module of the moments at the tool tip point according to the chip ejection angle and f growing (ap = 3.5 mm)

In Table 2 and Table 3 we also notice that $M_x$ is always positive whereas $M_{zCop}$ (resp. $M_{yCop}$) is always negative (resp. positive enough for ap = 1 Table 2). For ap (resp. f) growing there is a minimum about 50° (resp. 76°) for MzCop Figure 14 (resp. Figure 16). It is advisable to note that this minimum corresponds to a change of moments evolution; all the constituents of moments decrease until ap = 2 then grow from ap = 3.5. This is in agreement with the Figure 8. This analysis shows that the module of the moments (Fig. 12 and 13) is a function of $\theta^2$, ap and $f^2$. It is possible to write:

$$\text{IIMII} = \theta^2 ( k.ap + h.f^2 ) \tag{8}$$

where h and k are coefficients which depend on the couple tool/material. Other tests are necessary to obtain h and k.

We see that moments are more sensitive than force them to these parameters. Indeed if the strengths are always increasing with ap this is not any more the case with moments. The components of these pass by a minimum for ap = 2. On the other hand, when f varies the component Fy and the component Mx pass by a minimum for f=0.0625 (Table 3).

**Conclusion**

We brought to light a simple method to determine the center of the rigidity (rotation) dynamic of the system manufacturing directly from the measure of the complete mechanical torsor of the efforts applied at the tool tip point. It is an important element, which intervenes in a major way in the process of chip formation, which depends on it strongly.

We notice that there is a clear correlation between the evolution of the chip ejection angle and the evolution so many moments as central axes. However, the chip ejection angle seems to be more sensitive to the evolution of the moments than the resultant of the strengths led by the manufacturing. The linear evolutions (resp. parabolic) obtained enter the evolution of moments when ap (resp. f) increases according to the chip ejection angle is an important result for the simulations of current manufacturing. Besides evolution of the modules of the moments reported to θ when f (ap = 3.5 mm) and ap (f = 0.1 mm/rev) grow is a function of $\theta^2$, $f^2$, ap. We also notice that the angle of the central axes with the axis Z is parabolic type when f increases while it is linear when it is ap, which increases. These diverse results are important for the constitution of the simulation softwares of manufacturing outstanding discounted bills.

Other results of measures in the course of perusal also show the importance, which there is to follow the evolution of moments in the course of manufacturing, and will be presented in the next publication.

## Acknowledgements

The authors would like to thank the CNRS (UMR 5295) for the financial support to accomplish the project. This paper was supported by CNCSIS-UEFISCSU, project PNII-RU-code-194/2010.